# Experimental observations on interaction between a root and droplets in relation to aeroponic agriculture

Tejas Narasegowda[1,2*] and Navneet Kumar[3,4]

[1] *University Visvesvaraya College of Engineering, Bengaluru-560001; tejas.n@campusuvce.in*

[2] *Friedrich-Alexander-Universität Erlangen-Nürnberg, Germany; tejas.narasegowda@fau.de*

[3] *Indian Institute of Technology, Jammu-181221; navneet.kumar@iitjammu.ac.in*

[4] *Indian Institute of Science, Bengaluru-560012; navneetkumar@iisc.ac.in*

[*] *Correspondence: tejas.n@campusuvce.in*

**Abstract**

Aeroponics or Soil-less agriculture is a relatively new and recent type of practice, where plants are grown without soil while nutrient-rich water is provided via an atomized spray system to the suspended roots. Spray nozzles are easy-to-use in supplying water (and fertilizers) to (mainly) the roots and root hairs of the desired crop (or plant) for production. We characterize a spray nozzle delivering water vertically above against the gravity by measuring, experimentally, its (a) spray drift, (b) spray height, (c) maximum spray angle, (d) spray width, and (e) droplets sizes. Experiments were carried out at different inlet pressures and a majority of the above mentioned parameters were obtained by processing the images captured using optical (or high speed) camera, sometimes along a plane lighted by a high-power laser source. We also studied the spray (or jet) behaviour at different vertical heights and different horizontal planes using a unique polythene sponge method. We studied the mass flow rate, the absorption rate, and droplet size dynamics (as a function of time and pressure) using this method. The water drop/droplet interaction was also studied in the case of simpler porous and impervious surfaces as well. We believe that this study can be extrapolated to other nozzles (especially sprays) to obtain similar characteristic parameters. This study, hence, is critical in selecting the desired spray system for a given canopy and is also expected to be of some use in controlled agricultural practices such as in greenhouses and apartment rooms.

## 1. INTRODUCTION

Agriculture is at the root of our economic development. Open field agricultural practices, which have been commonly followed since ancient times, have proven insufficient for decades. The traditional way of agriculture is more dependent on the mercy of unpredictable nature, climate, weather, and seasons rather than technology. Still, we heavily depend on open field agriculture for our survival.

The world's population is expected to increase by 2 billion people in the next 30 years, from 7.7 billion currently to 9.7 billion in 2050 (by 25.97%) [1]. But the agricultural land is expected to come down by 17% in the next 30 years, from 36.6% currently to 19% in 2050 [2]. Zhu et al.,[3] in their recent study recorded that till 2014, farmland area has decreased at least in China. Hence, by 2050, the available agriculturally-fertile land may no longer be sufficient to support the increasing number of people depending on it for food. Therefore, it is important to find an alternative method for producing food in these circumstances. These methods are nowadays focused on precision agriculture needing skilled manpower amongst many other requirements. One of the solutions is 'Aeroponics'.

Some of the early works on the water-culture method for growing plants without soil were studied by Hoagland & Arnon [4], whose objective was to understand the fundamental factors which affect the plants' growth. In 1953, apple trees were grown outdoors with their roots in boxes where they were fed with a nutrient solution through spray [5]. A pioneering study was conducted on Aeroponic growth and its efficiency by Nir [6] and whose efficiency was further analyzed separately by Klotz [7] using a *Devilbiss* atomizer to provide a nutrient solution to the roots of the citrus plant. Stoner etal,.[8] showed that the Aeroponics system improved the root growth, survival rate, growth rate, and reduced maturation time. In 1988, The regeneration of plants using nutrient mist bio-reactor was devised by Weathers & Giles [9]. Mild-to-severe drought conditions, occurring naturally or man-made, motivated Hubick et al.,[10] to study and explore the Aeroponic agricultural system under a controlled environment. Massantini [11-13] described several types of Aeroponic systems as early as in the 1970's.

The other aspects like roots-based research such as root growth, root diseases, and antioxidant activity in the roots have also been investigated [14-17]. Further, the Aeroponic system was shown to measure the rate of nutrient uptake under varying conditions [18]. Later, the experimentations were extended to space flight applications for International Space Station which must be operated in a limited area [19,20]. Towards the application perspectives, mass production of Potato mini-tubers using aeroponics technique was achieved and argued to be more efficient and economical [21,22]. Some research works revealed that many medicinal plants cultivated in the aeroponic system are found to be the most significant for the production of quality biomass in a pesticide-free environment [23-25]. Some of the recent results show that, in tropical conditions, a quality seed can be produced using Aeroponic technology. The rate of seed multiplication and seed health can be improved by chilling the root zone [26]. Further, research on the aeroponics growth system was developed using an automated ultrasonic atomizer [27] with a monitoring system [28].

The most important component in the Aeroponics study is the Spray nozzle. In the Aeroponics spray system, the nutrients are sprayed at regular intervals. In 1963, Muras [29] explored the use of the nutrient spray in the culture of various vegetable species and discussed the role of spraying nutrients at finite targeted time intervals specific to a given plant. This work was followed by Shtrausberg [30], describing the significance of spraying nutrients mist at different intervals of time was analyzed for the tomato plant in the aeroponic chamber [31]. The spinning disk system of Zobel et al., [32] and "*Ein-gedi-system*" of Soffer [33] are the systems other than nozzles which also form a thin layer of water (nutrient solution) on the surface of the root. One of the recent studies explored the effect of plant growth with different aeroponic atomizers [34]. Significant effort has been expended in developing and evaluating the physics of atomization and characteristics of spray behaviour [35,36]. The universal adhesion model was proposed to predict the percentage adhesion of spray droplets impacting any leaf surfaces [37,38].

The main objective of this study was to establish correlations concerning parameters of nozzle characteristics and to select suitable nozzles for different vegetation culture. The other novelty of the present investigation is to understand the interaction between the sizes of the water drops/droplets with different surfaces such as plant roots and other porous materials like 3-D sponges and planar filter papers.

## 2. MATERIALS AND METHODS

Research protocol: The experiments were conducted in a chamber with cross-sectional dimensions of 625mm x 620 mm and a height of 1808mm. The bottom of the chamber consisted of a water tank which was used to deliver water to the nozzle through a booster pump. One of the four sides of the chamber was kept open for visualization purposes while the top was left open. The three closed sides of the chamber were covered with black cloth for imaging and filming purposes. The schematic of the complete experimental setup is seen in Figure 1. We used a Polyester sponge (94mm x 146mm x 64mm and mass of 6g) to obtain the amount of water absorbed at different locations. The sponge is held on an acrylic sheet that acts as a holder. We used a mobile weighing pan having the least count of 1g. This weighing pan was used to measure the sponge mass at regular intervals, thereby, yielding the absorbed amount of water. In this study, we used water for characterizing the nozzle since the percentage of fertilization is just 1-2% [39] and within this range, the flow characteristics is expected to remain similar to that of water. The main components used in the present experiments are seen in Figure 2.

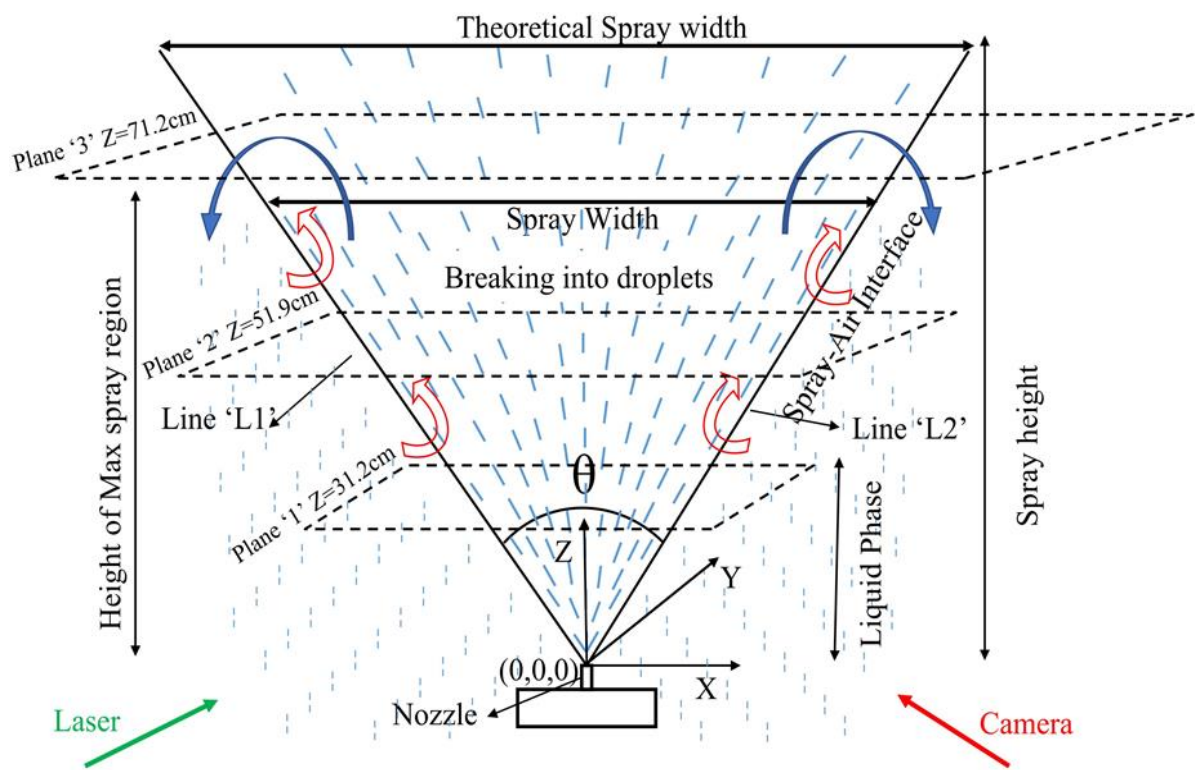

Figure 1: Schematic of the experiment setup.

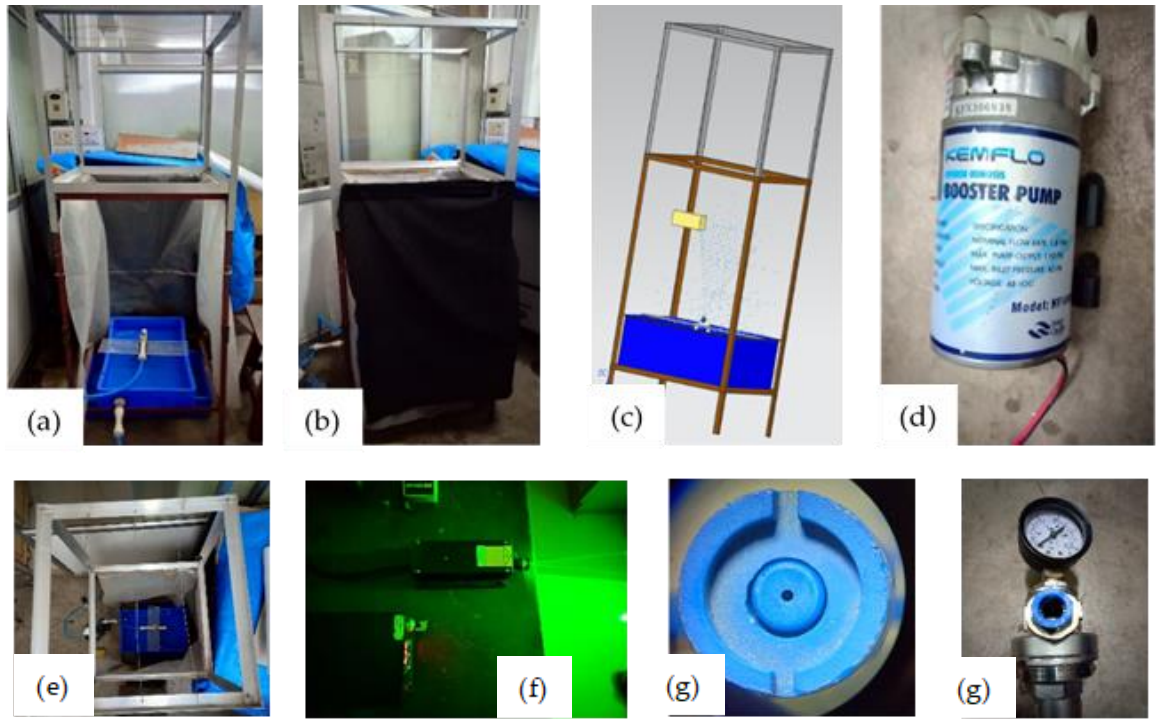

Figure 2: Components used in the present experimental study. (a) front view of the experimental setup, (b) side black covers, (c) schematic, (d) booster pump, (e) top view, (f) laser, (g) nozzle opening (0.60 mm diameter), and (h) pressure regulating va

### 2.1. Image Processing

Videos and images were captured using a combination of a high-power laser (512 nm) and sheet optics at different sections of the spray. Photographs and (slow-motion) videos were taken through a high-speed camera. The films and images were post-processed [36] in MATLAB which yielded the geometrical characteristics of the spray.

In the case of a horizontal surface, droplet-droplet and droplet-surface interactions are of three types; these are (1) bouncing-off of the droplets from the surface with either some retention or none, (2) film formation, and (3) collision between two or more droplets followed by the formation of a liquid film. (Figure 1)

For the inclined surfaces, droplet-surface interaction eventually leads to the formation of a thin film which (may) drip under the influence of the gravitational forces (commonly known as rivulets). In the case of the actual roots, any combination of the above-mentioned processes may occur.

## 3. Results and Discussion

### 3.1. Spray geometrical characteristics

The variations of the geometrical features of the spray with different input nozzle pressures are seen in Figure 3. The spray height achieved (see Figure 3a) increased at higher pressures while the spray width (defined as the lateral span of the spray, see Figure 3b) decreased considerably. However, it is the reduction in the spray angle (the angle at which the sprayed fluid fans out from the spray nozzle, see Figure 3c) that is the most prominent feature here.

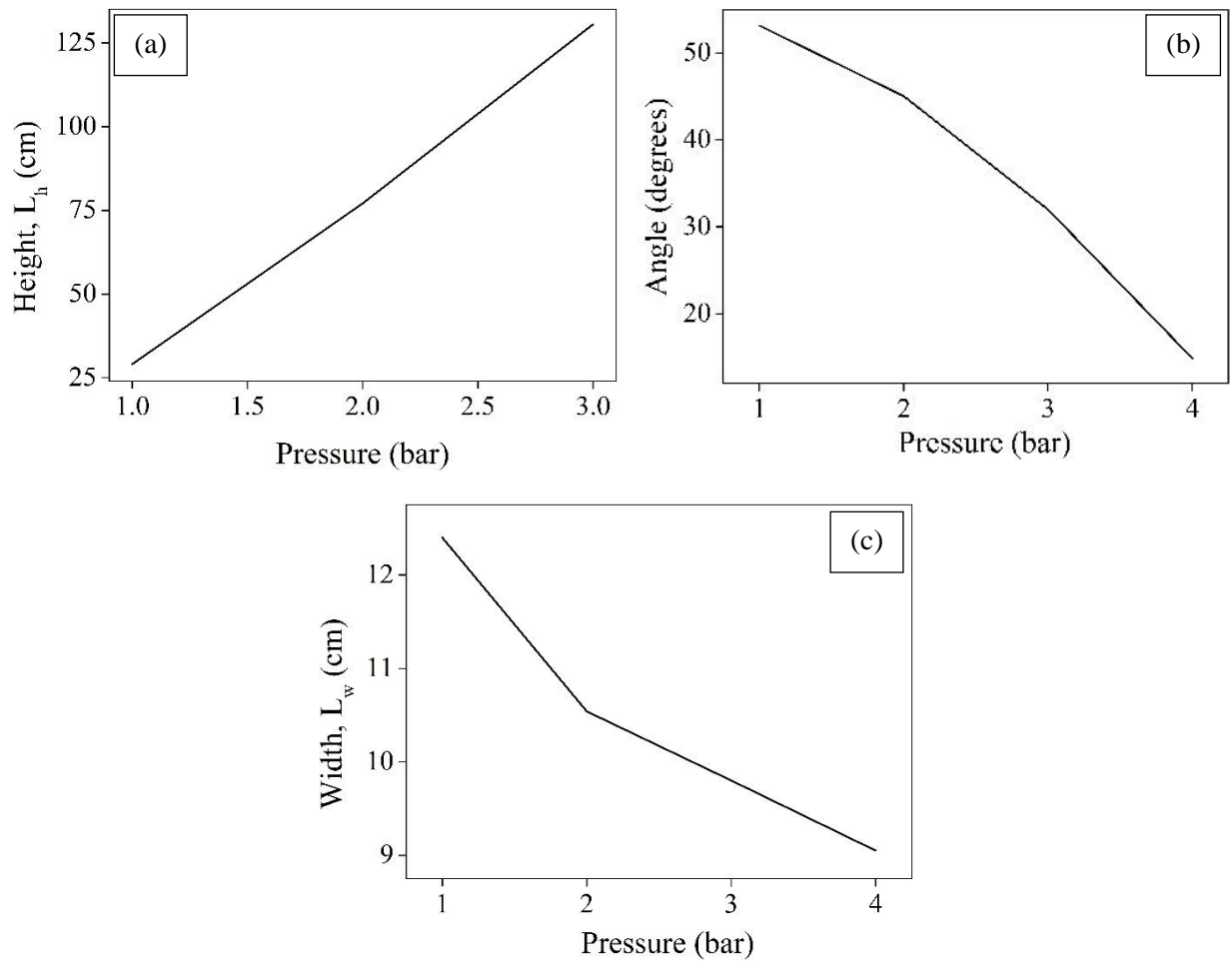

Figure 3: A plot of nozzle inlet pressure versus (a) spray height, (b) spray width, and (c) spray angle.

### *3.2. Spray interaction with a plant root and other model porous systems*

After only 2 seconds of the spray operation, the formation of a water film was observed enveloping the root; at 4 seconds, a series of drops fell from the root. These drops were considerably bigger in size and were observed to block the droplets' path. The water film must have formed (on the root) due to a large number of interactions between the droplets in a short duration of time. These interactions led to the formation of a stable water film (Figure 4a) which (a) may help in a continuous supply of water to the roots and (b) leads to water loss in the form of big water drops (see Figure S2 and Video S1)

A more careful study would be to characterize water absorption by counting "impinging water droplets" and "falling water drops" which would give the estimate of the "*Plant Water Uptake*". As seen here, the droplet-droplet and droplet-surface interactions are highly complicated and physically rich. In order to simplify this complex interaction, simple porous materials were also used; these are 3-D polymeric sponge and 2-D filter papers.

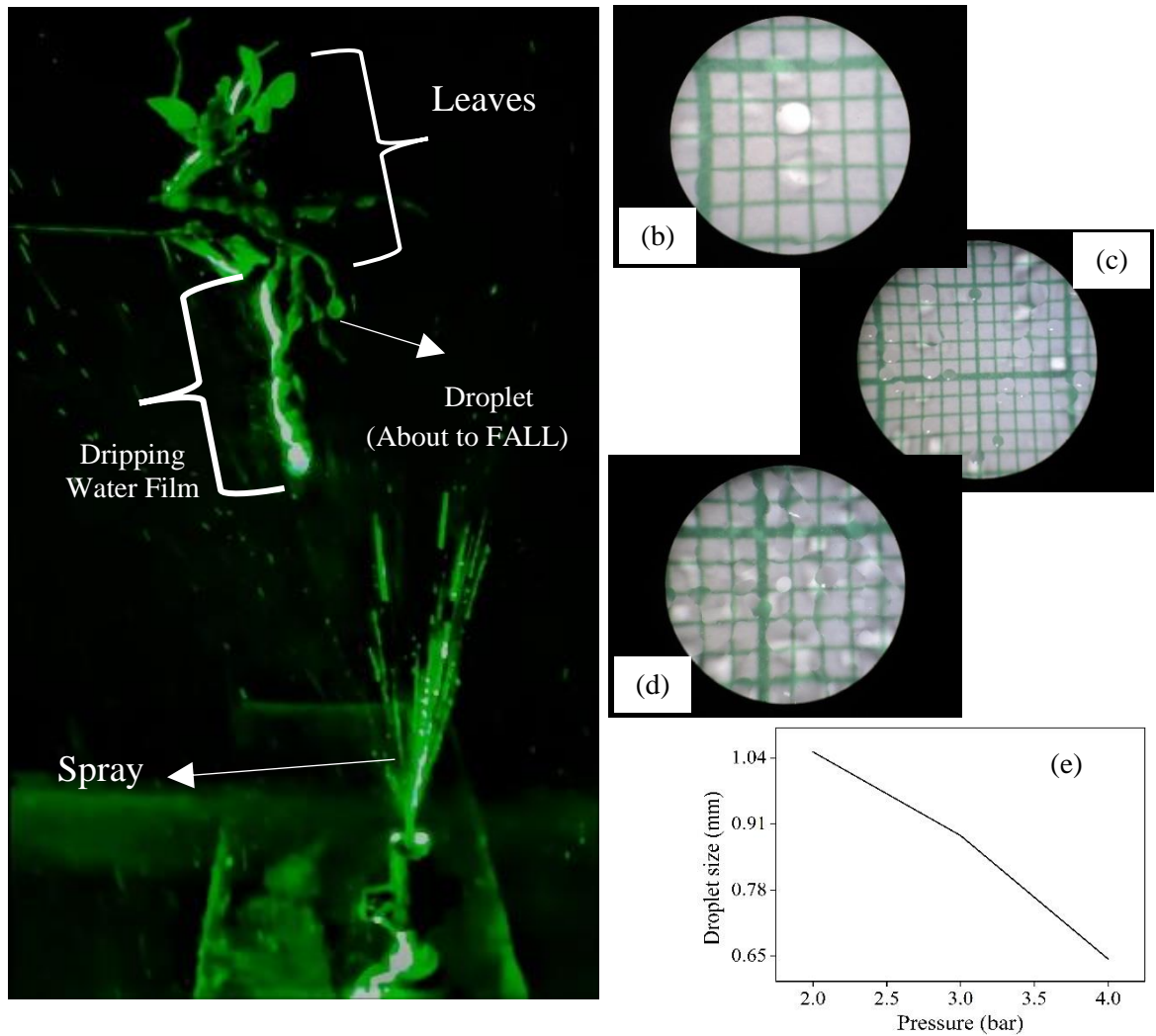

Figure 4: (a) Combination of droplet-droplet-root interaction in a spray cone. Accumulation followed by falling-off of a water drop is shown. Also seen is the formation of a water film around the plant root. Droplets resting on the glass plate at different pressures (b) 2, (c) 3, and (d) 4 bar respectively. Variation of the impacting droplet size with the nozzle inlet pressure is shown in (e).

It was observed that the sponges, when placed in the spray zone, initially gained weight by trapping the impacting water droplets for some time (this time varied between 4 minutes to 14 minutes depending on the sponge location). But after some time, they could not trap more liquid at which point a dripping film was observed in all the (24) experimental runs (Figure S5-S8). The formation of a liquid film in such porous systems seems to represent a steady state in terms of the amount of the liquid trapped. At larger heights, the sponge trapped less water and consumed more time to reach the steady state. Although the pore sizes in the sponge (Figure S9) were much higher than those in the roots, these results indicate that the impacting area (Figure S10) is an important parameter for consideration in an aeroponic system.

On the other hand, the pore sizes in the coffee filter paper were similar (of the order of 10 microns) to those in the roots. However, these papers were highly hydrophilic in nature and they were wetted completely within a few seconds. (Figure S11)

### 3.3. Spray interaction with a glass surface

Curvilinear complexities (in roots) were removed by using a flat and smooth glass plate to further study the droplet-surface interaction. This method indicated the impacting droplet size. One side of the glass plate was impinged with the spray while the other side was pasted with a graph sheet. The transparency of glass ensured the measurements of the collected/adhered droplets/drops under a microscope (Figure 4b-d).

The time of impact was a crucial parameter here. When the plate was placed for a long duration, a continuous film formed. Similar to the root case, here also, the film ended at the plate edge (see Video S2) and finally, falling drops were observed. Therefore, a long-time duration was not suitable to indicate the true droplet size striking the plate; these experiments were conducted for about 5 seconds only, sometimes even lesser. As seen in Figure 4(b-e), the droplet size reduced at higher pressure values. Droplet size and the impacting surface decide the fate of the droplet. Tiny droplets would tend to bounce back and hence are useless to the roots. On the other hand, a continuous water film seems to be a waste of water and nutrients but it would also aid in plant growth.

Experiments related to the Aeroponic systems were conducted with the primary focus on the characterization of the nozzle spray. The flow visualization was carried out using a planar laser sheet and high speed imaging of the water droplets emanating from a nozzle (dispensing the liquid in the vertically upward direction) and in some cases, the interaction between the water drops and root (and a few roots-like materials like 2-D filter paper and 3-D sponge) were also studied. The size of the water drops/droplets were obtained using a smooth glass plate kept in the spray zone for a few seconds.

Interactions between the droplets, coming out of the nozzle, and different surfaces were experimentally observed in order to mimic, to the first order, the behaviour of the plant roots in an aeroponic systems. For this purpose, real roots and some other root-like porous materials were used. The experimental observations clearly showed highly complicated interactions. The droplets impact the roots (and its other parts) and form a liquid film; this film led to continuous dripping of large drops at the root edges. However, the formation of a film may be required for the continuous supply of water and the nutrients in Aeroponic systems. Different types of droplet-surface interactions were, therefore, studied in order to understand root-drop coupling, a feature important in Aeroponic systems. With the glass plate and droplets system, we were able to identify the droplets sizes; this is an important parameter in connection to the aeroponic systems. These results open further opportunities in the relatively new agricultural practice. We expect this fundamental study to be useful in an aeroponic system, a new type of trend in Urban and precision agriculture.

**Conclusion**

Spray height linearly depends on the inlet pressure of the nozzle and the ratio of spray height and inlet pressure is constant. Spray width was drastically reduced at elevated inlet pressures. Spray angle decreased non-linearly with pressure. The sponge-mist interaction yielded that the amount of water absorbed reduces at locations away from the spray. Droplets-root interaction was shown to be extremely complicated with multiple phenomena occurring simultaneously. The droplets impact the roots (and its other parts) and form a liquid film, which led to continuous dripping of large drops. This type of interaction was also seen in the model porous systems, 3-D sponges, and 2-D filter papers. Similar to the root case, dripping of drops were seen in the case of the sponges.

Therefore, the different types of droplet-surface interactions were studied in order to understand the root-drop interaction, a feature important in the Aeroponic systems.

**Acknowledgement**

We acknowledge the financial support from the Ministry of Earth Sciences (India) under the grant MESO/0034 and Indian Council of Agricultural Research (ICAR) under the grant ICAR-0014. We appreciate the help, apparatuses, and funding provided by Prof. Jaywant H Arakeri (IISc Bangalore). We also thank Mr. C. Dharuman (Senior Scientific Officer, IISc Bangalore) and Mr. Rakesh (Bachelors' graduate from UVCE, Bangalore) for their help during the experiments.

**Supporting Information**

Videos and data required to reproduce this research are available at https://doi.org/10.6084/m9.figshare.12115731.